\documentclass[10pt,journal,twoside,final]{IEEEtran}

\usepackage{graphicx}

\usepackage{amsmath}
\usepackage{cite}
\usepackage{multirow}
\usepackage{url}
\usepackage{bigstrut}
\usepackage{array}
\newcolumntype{L}[1]{>{\raggedright\let\newline\\\arraybackslash\hspace{0pt}}m{#1}}
\newcolumntype{C}[1]{>{\centering\let\newline\\\arraybackslash\hspace{0pt}}m{#1}}
\newcolumntype{R}[1]{>{\raggedleft\let\newline\\\arraybackslash\hspace{0pt}}m{#1}}

\ifCLASSOPTIONcompsoc
\usepackage[caption=false,font=normalsize,labelfont=sf,textfont=sf]{subfig}
\else
\usepackage[caption=false,font=footnotesize]{subfig}
\fi

\usepackage{algorithm}
\usepackage{algpseudocode}
\algnewcommand\algorithmicinput{\textbf{INPUT:}}
\algnewcommand\INPUT{\item[\algorithmicinput]}
\algnewcommand{\algorithmicoutput}{\textbf{OUTPUT:}}
\algnewcommand\OUTPUT{\item[\algorithmicoutput]}
\algrenewcommand{\algorithmiccomment}[1]{\hskip3em$\%$ #1}

\usepackage[switch]{lineno}

\newcommand*\patchAmsMathEnvironmentForLineno[1]{%
\expandafter\let\csname old#1\expandafter\endcsname\csname #1\endcsname
\expandafter\let\csname oldend#1\expandafter\endcsname\csname end#1\endcsname
\renewenvironment{#1}%
{\linenomath\csname old#1\endcsname}%
{\csname oldend#1\endcsname\endlinenomath}}%
\newcommand*\patchBothAmsMathEnvironmentsForLineno[1]{%
\patchAmsMathEnvironmentForLineno{#1}%
\patchAmsMathEnvironmentForLineno{#1*}}%
\AtBeginDocument{%
\patchBothAmsMathEnvironmentsForLineno{equation}%
\patchBothAmsMathEnvironmentsForLineno{align}%
\patchBothAmsMathEnvironmentsForLineno{flalign}%
\patchBothAmsMathEnvironmentsForLineno{alignat}%
\patchBothAmsMathEnvironmentsForLineno{gather}%
\patchBothAmsMathEnvironmentsForLineno{multline}%
}

\begin{document}

\title{Channel-Envelope Differencing Eliminates Secret Key Correlation: LoRa-Based Key Generation in Low Power Wide Area Networks}

\author{Junqing~Zhang,
Alan~Marshall,~\IEEEmembership{Senior Member,~IEEE},
and Lajos~Hanzo,~\IEEEmembership{Fellow,~IEEE}
\thanks{J. Zhang and A. Marshall are with the Department of Electrical Engineering and Electronics,  University of Liverpool, Liverpool, L69 3GJ, U.K. (emails: Junqing.Zhang@liverpool.ac.uk; Alan.Marshall@liverpool.ac.uk)
}
\thanks{L. Hanzo is with the School of ECS, University of Southampton, Southampton SO17 1BJ, U.K. (email: lh@ecs.soton.ac.uk) L. Hanzo would like to gratefully acknowledge the ERC's financial support of his Advanced Fellow Grant.}
}

\maketitle

\begin{abstract}
This paper presents automatic key generation for long-range wireless communications in low power wide area networks (LPWANs), employing LoRa as a case study. Differential quantization is adopted to extract a high level of randomness. Experiments conducted both in an outdoor urban environment and in an indoor environment demonstrate that this key generation technique is applicable for LPWANs, and shows that it is able to reliably generate secure keys.
\end{abstract}
\begin{IEEEkeywords}
Internet of Things, low power wide area networks, physical layer security, key generation, LoRa/LoRaWAN
\end{IEEEkeywords}

\section{Introduction}
The Internet of Things (IoT) is capable of connecting people, things, and the environment. This revolution heavily relies on secure data communications, which are currently maintained by classic cryptographic algorithms and protocols. In particular, public key cryptography (PKC) has been the de facto scheme for distributing keys to the users in modern communication and computer networks. However, its application in the IoT remains a challenge owing to the limited computational and battery capacity, as well as the requirement of a public key infrastructure for distributing the public keys.

Key generation from the wireless channel between any pair of users has become a promising design alternative to complement PKC. The keys generated can be used for the symmetric encryption schemes in different layers of the protocol stack, e.g., the Wi-Fi Protected Access (WPA) for the Wi-Fi MAC layer encryption or for Transport Layer Security (TLS) in the transport layer.
It is particularly for protecting IoT systems that contain large numbers of resource-limited devices~\cite{zhang2016review}. A comparison of resource and energy consumption between the key generation and  elliptic curve-based Diffie-Hellman (ECDH) procedure, which is a popular PKC scheme, has been carried out in~\cite{zenger2016authenticated}. Key generation has been demonstrated to be more cost-efficient. Explicitly, ECDH consumes 98 times more energy and imposes 1289 times higher complexity than key generation, when both are implemented by an 8-bit Intel MCS-51 micro-controller~\cite{zenger2016authenticated}. In addition, key generation does not require any assistance from a third party, which is suitable for many decentralized or device-to-device IoT applications.

The received signal strength indicator (RSSI) has been the most popular parameter because of its wide availability in the transceivers and network interface cards. This has been evidenced by its wide applications in Wi-Fi~\cite{jana2009effectiveness,zan2013key,zhang2016experimental,zhang2016access}, ZigBee~\cite{gungor2015secret,zenger2016passive} and Bluetooth~\cite{premnath2014secret}, etc. However, all these wireless techniques only support operations in short-range environments, typically within 100 meters. The channel may be deemed reciprocal in such environments. For example, we carried out key generation for Wi-Fi in an indoor office scenario~\cite{zhang2016experimental}. The RSSI varied from -50~dBm to -25~dBm and random keys were generated from the reciprocal measurements.

In reality, many IoT applications operate in longer-range environments, e.g., vehicular communications. 
There have been several long range standards designed for low-power wide area networks (LPWANs), including LoRa, Narrowband IoT, and Sigfox, etc.
A very recent conference contribution applies key generation with LoRa~\cite{ruotsalainen2018towards}, but the experiments are carried out in short-range environments, since the received power only has a 20~dBm variation.
In contrast to short-range wireless communications, the channel conditions in long-range networks may vary significantly due to the shadowing of buildings in urban environments.

This paper investigates the key generation in LPWANs with long range communications, by employing LoRa as a case study. Our work observed large RSSI variations of the devices, and used differential quantization to extract the channel's randomness. Experiments have been carried out both in an outdoor urban environment and in an indoor environment. The system is shown to exhibit beneficial channel reciprocity as confirmed in terms of cross-correlation and key disagreement ratio (KDR), and a sufficiently high degree of randomness.

\section{Overview of LoRa/LoRaWAN}
LoRa is a  physical layer modulation technique developed by Semtech while LoRaWAN is the MAC protocol maintained by the LoRa Alliance~\cite{LoRaWAN}. This section briefly introduces the relevant background and a detailed introduction can be found in~\cite{augustin2016study}. 

\subsection{Physical Layer}
LoRa uses chirp spread spectrum (CSS) modulation, which is immune to multipath and Doppler shift. It is quite robust and achieves a receive sensitivity as low as -148~dBm, which is eminently suitable for long range communications.
The main parameters include  bandwidth, spreading factor and code rate, which can be adjusted according to the specific requirements of sensitivity, communication range, and data rate. 

\subsection{MAC Layer}
LoRaWAN relies on a star network topology involving gateways and end devices. 
According to the different configurations of the receive windows at the end device, there are three device types, namely Class A, B, and C. The Class A functionality is mandatory, which is explained in this paper.

The Class A end device can initiate the uplink transmission. It will then open two receive windows after a certain delay. 
In other words, the gateway can only send a downlink frame to the end device, provided that it receives an uplink frame. The power consumption of the end device is thus kept very low. 
LoRaWAN also defines the so-called confirmed data message type, which must be acknowledged by the receiver. The confirmed data message and its ACK message constitute a pair of bidirectional transmissions, which can be leveraged for key generation.

\subsection{LoRaWAN Security Mechanism}
LoRaWAN has a rigorous security mechanism for protecting both the application payload and the communication sessions. The encryption algorithm is based on the one used in IEEE~802.15.4, which employs advanced encryption standard (AES) with a key length of 128~bits. 
LoRaWAN defines two activation methods, namely activation by personalization (ABP) and over-the-air activation (OTAA). In the ABP, the session keys are programmed into the end devices during manufacturing, which cannot be updated. In the OTAA, the session keys are generated from the device's root keys, including AppKey and NwkKey. 
However, similar to other symmetric encryption schemes, the distribution technique of the device's root keys is not defined in the standard.
Inspired by this, we will propose an innovative key generation scheme by exploiting the unpredictable features of the wireless channel between any pair of devices.

\section{Key Generation Protocol}\label{sec:protocol}
A full key generation protocol usually includes channel probing, quantization, information reconciliation, and privacy amplification~\cite{zhang2016review}. A pair of legitimate users,  Alice and Bob, will carry out channel probing by performing bidirectional channel measurements. Once sufficient results are collected, they will separately convert the analog measurements into binary sequences using a quantizer. Since there may be mismatch between the keys at Alice and Bob due to  noise and asynchronous sampling, information reconciliation is adopted for allowing them to agree on the same keys. Finally, privacy amplification is employed to remove the information leakage. These four steps will be discussed in detail as follows.

\begin{figure}[!t]
	\centering
		\includegraphics[width=3.1in]{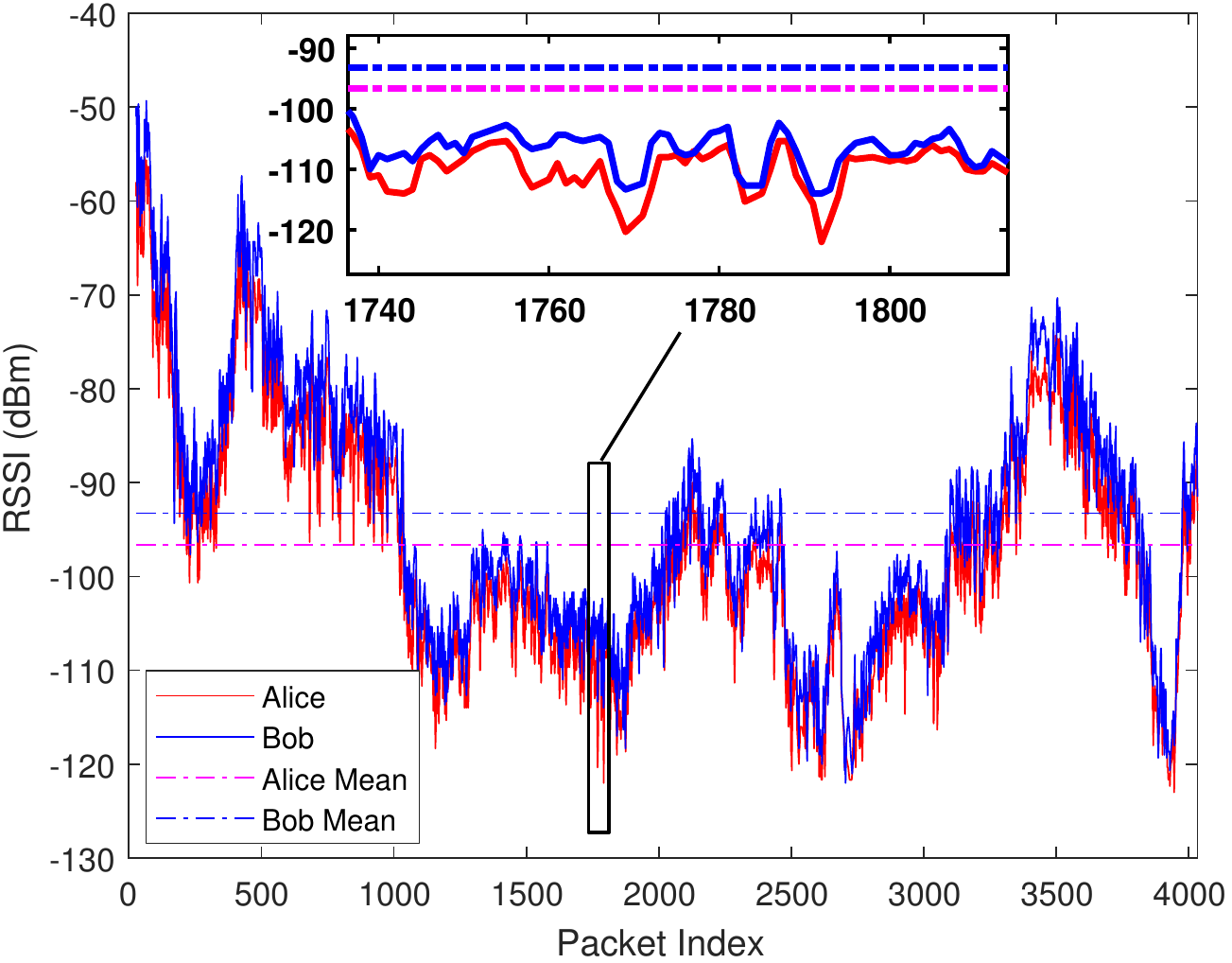}
	\caption{RSSI of Alice and Bob of experiment carried out within the campus of University of Liverpool.}
	\label{fig:campus}
\end{figure}

Channel probing harvests the randomness from the wireless channel.
During the $i^{th}$ probe, Alice sends a packet to Bob who will measure the RSSI $X_B(i)$. Upon receiving it, Bob will reply a message to Alice, who will measure the RSSI $X_A(i)$.  Alice and Bob will keep these bidirectional transmissions until they collect sufficient data. 
An example of the RSSI of Alice and Bob collected from an outdoor experiment is shown in Fig.~\ref{fig:campus}, and the detailed setup will be discussed in Section~\ref{sec:exp}. It is worth noting that RSSI measurements can be carried out during regular data transmissions and no dedicated packet exchange will be required.

Quantization in key generation discretizes the analog measurements into a binary sequence, which works in a similar manner to the classic analog-to-digital converter (ADC). Absolute value-based quantization is commonly used for comparing   measurements to thresholds and then assigning binary values to the outcome.
For example, the mean value-based quantization will assign a 1 to any data above the mean value and a 0 to any data below the mean value.
However, the RSSI output of Fig.~\ref{fig:campus} varies from -123~dBm to -49~dBm, which is quite a large variation. There are many consecutive samples above/below the mean value. Hence, the mean value-based quantizer will  result in long runs of continuous 1s and 0s, which will not be random at all. Owing to this impediment, it is not adopted in this paper.
This scheme may be improved by first partitioning the measurements into smaller blocks and then quantizing each block separately, as in the adaptive secret bit generation of~\cite{jana2009effectiveness}. However, due to the large variation of RSSI output in LoRa measurements, it is challenging to determine the block size. 

We propose to carry out the quantization based on the differential value, namely the difference between  adjacent values, as shown in Algorithm~\ref{alg:differential}. The differential quantization concept was originally proposed in~\cite{zan2013key}. For each user $u$, $u = \{A, B\}$, it will carry out the quantization separately. Whenever a new RSSI, $X_u(i+1)$, is measured, the user $u$ will compare it to the previous one, $X_u(i)$, and assign $K_u(i)$ as 1/0, when it is larger/smaller than the previous RSSI. 
The RSSI measurement may not be very accurate because of using low cost hardware, hence the RSSI resolution $\epsilon$ is introduced. 
The RSSI values having variation smaller than $\epsilon$ are thus dropped in order to improve the robustness against the measurement imperfection.  As each packet has a unique packet sequence index,
the index of the dropped RSSI values is shared between Alice and Bob, so that they can maintain a common index.

Compared to the absolute value-based quantization, differential quantization captures the relative changes of the RSSI values and the channel conditions. This is beneficial because it produces the key bits based on the comparison between the adjacent measurements, which does not require any adaptive adjustment based on the channel conditions. It is therefore much more lightweight for implementation.
\begin{algorithm}[!t]
\caption{Differential-based quantization algorithm}
\begin{algorithmic}[1]
\INPUT{$X_{u}$ \Comment{RSSI of user $u$}}
\INPUT{$\epsilon$ \Comment{RSSI resolution}}
\OUTPUT{$K_{u}$ \Comment{Generated key sequence of user $u$}}

\For{$i \gets 1 \:\:\:\: \textbf{to} \:\:\:\:  N-1$}
\If{$X_{u}(i+1) > X_{u}(i) + \epsilon$}
    \State{$K_{u}(i) = 1$}
\ElsIf{$X_{u}(i+1) < X_{u}(i) - \epsilon$}
    \State{$K_{u}(i) = 0$}
\Else		
\State{$X_{u}(i)$ dropped}
\EndIf
		
\EndFor
\end{algorithmic}
\label{alg:differential}
\end{algorithm}

Alice and Bob will respectively produce $K_A$ and $K_B$  after quantization. However, as shown in Fig.~\ref{fig:campus}, the channel measurements $X_A$ and $X_B$ are not identical, which results in disagreement between $K_A$ and $K_B$. Information reconciliation is thus employed to correct the disagreement. Secure sketch is one of the popular protocols~\cite{dodis2008fuzzy}, as shown in Algorithm~\ref{alg:ss}. It exploits the correction capability of  error correction codes (ECCs)~\cite{hanzo2002turbo}, e.g., BCH, LDPC, etc. The ECC has a maximum error correction capability of $t$ errors. When the key disagreement, quantified by the Hamming distance, is lower than $t$, it can be corrected. Finally, because there is information exchanged publicly during the information reconciliation, privacy amplification, e.g., by employing hash function, is used to remove the information leakage.
\begin{algorithm}[!t]
\caption{Information reconciliation - secure sketch}
\begin{algorithmic}[1]
\INPUT $K_A$, $K_B$ \Comment{Quantized keys of Alice and Bob}
\INPUT $C$ \Comment{ECC set shared by Alice and Bob}
\OUTPUT $K_A$, $K_{B'}$ \Comment{Reconciled key}
\State{Alice randomly selects a code $c$ from an ECC set $C$ }
\State{Alice calculates $s = \text{XOR}(K_A,c)$}
\State{Alice transmits $s$ to Bob through a public channel}
\State{Bob receives $s$}
\State{Bob calculates $c_B = \text{XOR}(K_B,s)$}
\State{Bob decodes $c_B$} to get $c$ \Comment{When $dis(c-c_B) < t$}
\State{Bob calculates $K_{B'} = \text{XOR}(c,s) = K_A$}~\Comment{Alice and Bob agree on the same key}
\end{algorithmic}
\label{alg:ss}
\end{algorithm}

\section{Experimental Evaluation}\label{sec:exp}
\subsection{Setup}
A testbed was built using Arduino Uno and LoRa/GPS Shield that uses Semtech SX1276 as the LoRa transceiver. 
The RadioHead library~\cite{RadioHead} is used, which provides the function to obtain the packet's RSSI. RSSI has been used extensively in key generation to represent the link quality, and is also used in this paper.
Two LoRa modules, termed as Alice and Bob, are configured with the same parameters, including carrier frequency of 868.1~MHz, bandwidth of 125~kHz, transmission power of 13~dBm and spreading factor of 7. These two modules will carry out bidirectional channel measurements, as introduced in Section~\ref{sec:protocol}. The RSSI values are transferred to the PC via a serial port and further processed by Matlab.

Even when there is no channel variation or interference, the received power may fluctuate because of the imperfect hardware characteristics.
In order to quantify the resolution of RSSI, we carried out a calibration experiment in an anechoic chamber at the University of Liverpool. As shown in Fig.~\ref{fig:anechoic_chamber}, two LoRa devices were placed about two meters apart, which is a totally static and line-of-sight (LoS) scenario with no interference from other networks. The experiment ran for about 15 minutes and collected 4000 packets at each side. The RSSI of Alice and Bob is shown in Fig.~\ref{fig:anechoic_chamber}. As can be observed, while there are some spikes in Alice's RSSI, most of the RSSI values of Alice and Bob only have a 2 dBm variation. Therefore, we set $\epsilon = 2$ for the differential quantization.
\begin{figure}[!t]
	\centering
		\includegraphics[width=2.2in]{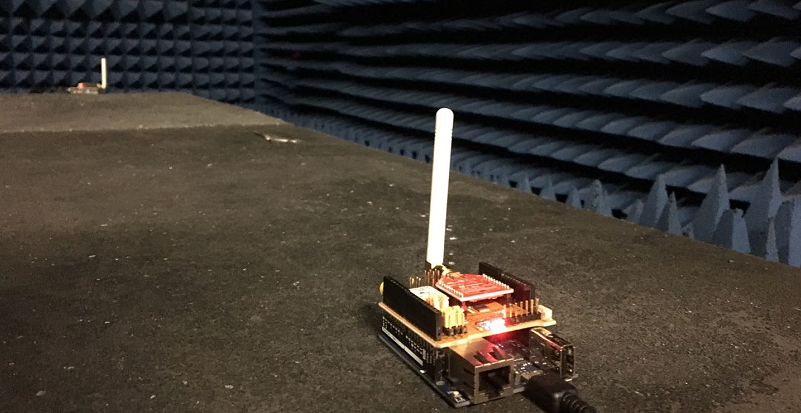}
	\caption{The placement of Alice and Bob in the anechoic chamber, University of Liverpool.}
	\label{fig:anechoic_chamber}
\end{figure}

\begin{figure}[!t]
	\centering
		\includegraphics[width=3.1in]{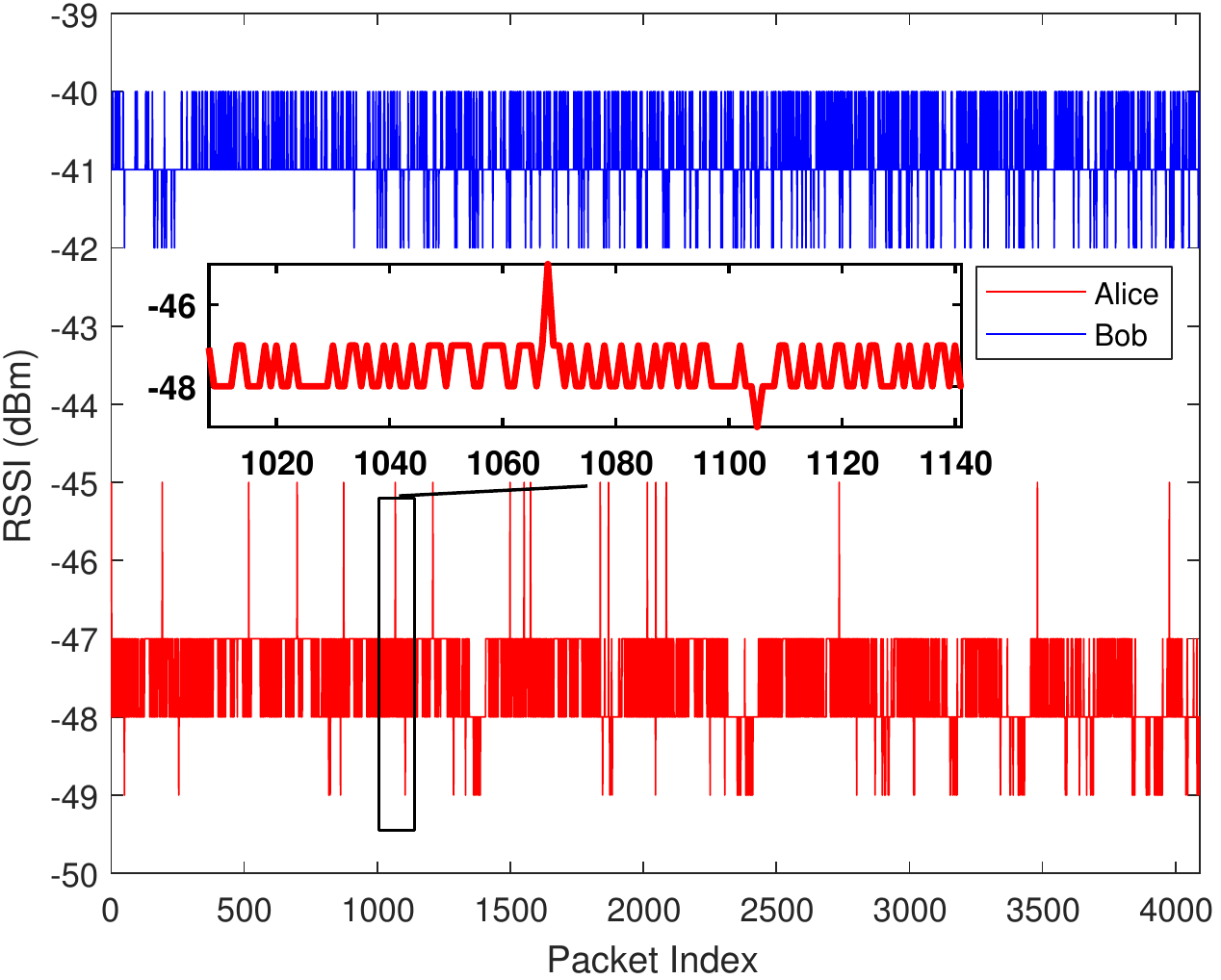}
	\caption{RSSI of Alice and Bob of experiment carried out in the anechoic chamber, University of Liverpool.}
	\label{fig:anechoic_chamber}
\end{figure}

\begin{figure}[!t]
	\centering
		\includegraphics[width=1.8in]{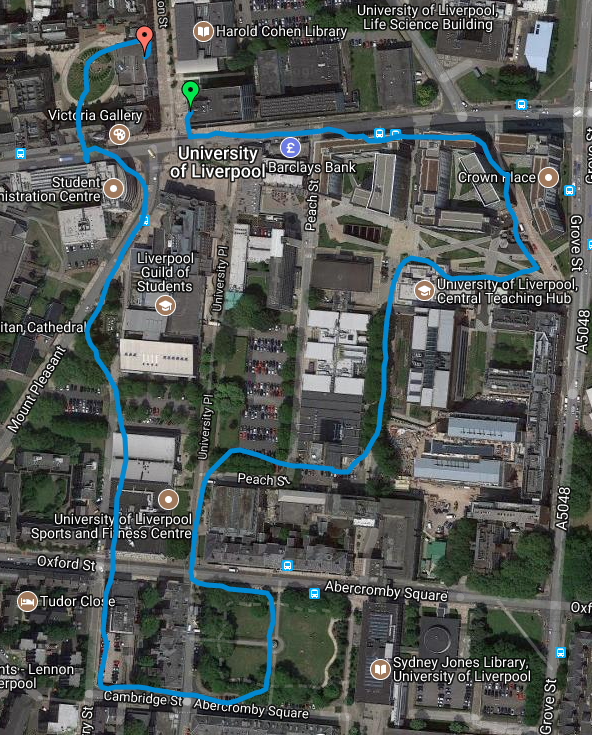}
	\caption{The trajectory of Bob in the campus of University of Liverpool.}
	\label{fig:route}
\end{figure}

We then carried out two tests.
Alice was placed in an indoor office of the second floor of the Department of Electrical Engineering and Electronics building (EEE), University of Liverpool, the green point in Fig.~\ref{fig:route}. 
In the outdoor experiment, Bob was moving at a walking speed, i.e., about 2 meters per second, in the campus of University of Liverpool. Bob moved from the green point to the red point with the detailed trajectory shown in Fig.~\ref{fig:route}. This is a typical urban environment with many buildings causing severe path loss and shadowing. The farthest distance between Alice and Bob in the experiment was about 500 meters. The experiment lasted 21 minutes and collected about 4000 packets in total.
In the second (indoor) experiment, Bob was moving inside the six-storey EEE Department building up and down. This is a typical indoor environment with rich multipath. The indoor experiment lasted 10 minutes and collected about 2300 packets.

\subsection{Results}
\begin{figure}[!t]
	\centering
		\includegraphics[width=3.1in]{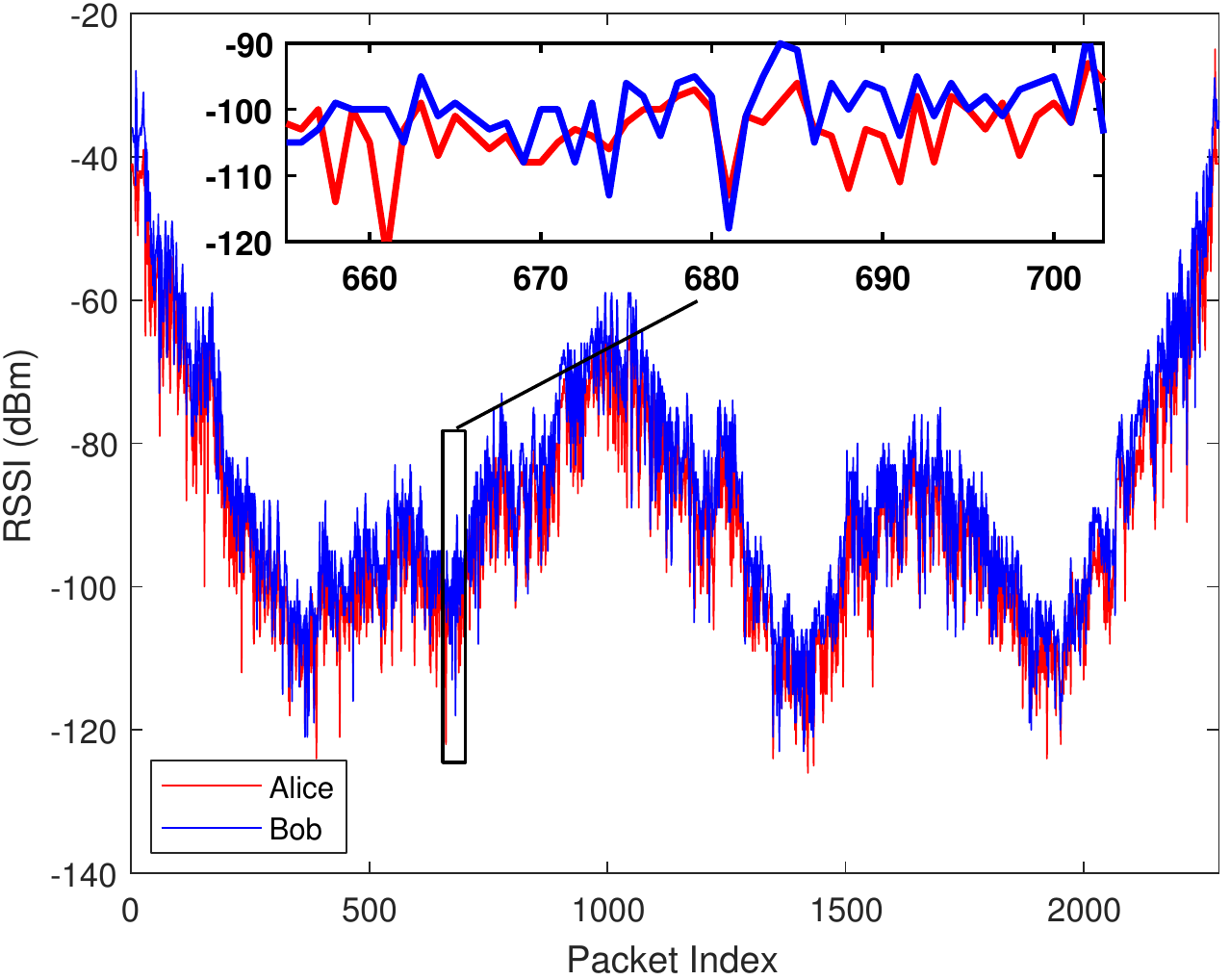}
	\caption{RSSI of Alice and Bob of experiment carried out inside the building of EEE Department, University of Liverpool.}
	\label{fig:indoor}
\end{figure}

The RSSI of the outdoor urban and indoor experiments is shown in Fig.~\ref{fig:campus} and Fig.~\ref{fig:indoor}, respectively. Both have very large variations. We use Pearson's cross-correlation coefficient and KDR to characterize the channel reciprocity, and randomness test to evaluate the quality of the key sequence~\cite{zhang2016review}. Cross-correlation describes the similarity between any two signals while KDR measures the ratio of different bits between two sequences. Randomness of the key determines the security level because a non-random key will be subject to brute force attacks.

Pearson's cross-correlation coefficient is defined as
\begin{align}
	\rho = \frac{\sum_{i=1}^{N}(X_A(i) - \mu_{X_A})(X_B(i) - \mu_{X_B})}{\sqrt{\sum_{i=1}^{N}(X_A(i) - \mu_{X_A})^2}\sqrt{\sum_{i=1}^{N}(X_B(i) - \mu_{X_B})^2}},
\end{align}
where $\mu_{X_u}$ is the mean value of $X_u$. 
The KDR is defined as
\begin{align}
	\textit{KDR} = \frac{\sum_{i = 1}^{l_k}|K_A(i)-K_B(i)|}{l_k},
	\label{eq:KDR}
\end{align}
where $l_k$ is the length of the key sequence. 
The correlation coefficients in the outdoor and indoor experiments are 0.9582 and 0.9689, respectively.
The correlation coefficients are high, which indicates a high grade of reciprocity.
The KDRs in the outdoor and indoor experiments are 0.0529 and 0.0399, respectively.
The KDR is quite low when differential quantization is used. As we analyzed in our previous work~\cite{zhang2016efficient}, a BCH $(n,k,t)$ code can correct $t/n$ mismatch, e.g., a BCH (15,3,3) can correct 20\% mismatch.
The KDR in this paper is thus well within a classic code's correction capability.

The National Institute of Standards and Technology (NIST) randomness test suite is the most popular tool for evaluating the randomness of the true/pseudo random number generator~\bstctlcite{IEEEexample:BSTcontrol}\cite{rukhin2010statistical}. It has been widely applied in  key generation research and it is also adopted in this paper for evaluating the randomness of the key sequence generated. Each test will return a p-value and when this is bigger than a threshold, e.g., 0.01 in this paper, the sequence passes the test. The randomness test results of the quantized key sequence is shown in Table~\ref{tab:random_test} and the key sequence passes all the tests.
\begin{table}[!t]
  \centering
  \caption{Randomness Test Results}
\begin{tabular}{|l|l|l|}
\hline
      & \multicolumn{1}{l|}{Outdoor} & \multicolumn{1}{l|}{Indoor} \\
\hline
Sequence Length & 397   & 376 \\
\hline
Frequency & 0.515 & 0.537 \\
\hline
Block frequency & 0.905 & 0.677 \\
\hline
Runs & 0.023 & 0.343 \\
\hline
Longest run of 1s & 0.887 & 0.331 \\
\hline
DFT & 0.792 & 0.85 \\
\hline
Serial & 0.3, 0.76   & 0.048, 0.048  \\ 
\hline
Appro. entropy & 0.065 & 0.1 \\
\hline
Cum. sums (fwd) & 0.538 & 0.432 \\
\hline
Cum. sums (rev) & 0.896 & 0.919 \\
\hline
\end{tabular}%
  \label{tab:random_test}%
\end{table}%

\section{Key Generation with LoRaWAN}
We have applied key generation relying on LoRa in the previous sections. While we can exploit the transmissions between the LoRaWAN gateway and end devices for key generation, there are special features and configurations in LoRaWAN, which require further careful considerations. 

The LoRaWAN standard supports up to 16 channels in total.
For example, The Things Network, a global IoT network hosting thousands of LoRaWAN gateways, defines eight frequencies~\cite{LoRaWANFrequency}.
In order to decrease the interference, the end device changes the carrier frequency in a pseudo-random fashion for every transmission. This pseudo-random frequency hopping is detrimental to key generation, because the channel conditions of different frequencies are not reciprocal. 

Fortunately, LoRaWAN also specifies that the downlink ACK frame should be at the same frequency as that of the corresponding uplink data frame~\cite{LoRaWAN}. 
The end device first transmits an uplink confirmed data packet to the gateway. Upon receiving the data packet, the gateway will respond with a downlink ACK frame. The carrier frequencies of these two packets should be the same. Therefore, the key generation probing process can be carried out using the confirmed uplink data frame and downlink ACK frame pairs.

\section{Discussion}
The keys generated can be used for any cryptographic schemes, which require a symmetric key. These schemes do not require a fast key update rate. For example, Wi-Fi recommends changing the keys, a 128-bit binary sequence, every hour. In our scheme, each differential comparison will produce one bit. Hence the key generation rate is  less than or equal to 1 bit per measurement, which should be sufficiently fast to generate keys at the required rate.

Key generation requires a dynamically fluctuating channel in order to produce random keys. When  users are stationary, the channel variation is introduced by  objects moving in the environment. Even in a totally static environment, multiple antennas and frequency diversity can be exploited~\cite{huang2013fast}. However, in many of the smart city and intelligent transportation systems, user movements and channel variations are indeed present as a much needed source of channel randomness.

Key generation is subject to passive eavesdropping~\cite{zenger2016passive,zhang2016access}, where the eavesdroppers record all the transmissions and try to crack the system. According to communication theory, when the eavesdropper is located sufficiently far from the legitimate users, the eavesdropping channel is uncorrelated with the legitimate channel. 
The seminal work in~\cite{ahlswede1993common,maurer1993secret} has laid the information-theoretical foundation for key generation, which formulates 
 the secret key capacity as
\begin{align}
	 C_{sk} \geq  I(X_A;X_B) - \min[I(X_A;X_E),I(X_B;X_E)],
\end{align}
where $X_A$, $X_B$, and $X_E$ are the observations of Alice, Bob, and Eve, respectively. When $ C_{sk}>0$, key generation can be carried out securely.
Key generation security under passive eavesdropping has been validated in~\cite{zenger2016passive} through extensive ZigBee-based experiments. In particular, the work in~\cite{ruotsalainen2018towards} demonstrates that  legitimate users can still generate keys securely by using LoRa, when the eavesdropper is located only 0.15~m or 2~m away.

\section{Conclusions and Future Work}\label{conclusions} 
This paper applies key generation with the LoRa technology and investigates automatic key generation performance in a long-range environment. Because of the large variation of the channel conditions and RSSI values, we employed differential quantization to extract the channel's randomness.
We carried out experiments both in an outdoor urban environment and in an indoor environment to demonstrate that LoRa-based key generation has a good performance. Key generation application in LoRaWAN was also discussed and was shown to be feasible by leveraging the uplink confirmed data packet and downlink ACK packet.



\end{document}